\documentstyle[epsfig]{aipproc}

\begin{document}
\title{COMPTEL Observations of the Blazars 3C~454.3 and CTA~102}

\author{S.~Zhang$^{1,5}$, 
W.~Collmar$^{1}$, V.~Sch\"onfelder$^{1}$, 
H.~Bloemen$^{2}$, W.~Hermsen$^{2}$, 
M.~McConnell$^{3}$, K.Bennett$^{4}$, O.R.~Williams$^{4}$}
\address{
$^{1}$Max-Planck-Institut f\"ur extraterrestrische Physik, Garching,
Germany \\
$^{2}$Space Research Organization Netherlands, Utrecht, The Netherlands \\
$^{3}$Space Science Center, University of New Hampshire, Durham, USA\\
$^{4}$Astrophysics Division, ESTEC, Noordwijk, The Netherlands \\
$^{5}$High Energy Astrophysics Lab, IHEP, P.O.Box 918-3, Beijing, China
  }

\maketitle

\vspace{-0.3cm}
\begin{abstract}
We have analyzed the two blazars of 3C~454.3 and CTA~102  using all available  
COMPTEL data from 1991 to 1999.
In the 10-30 MeV band, emission from the general direction of the sources is found at the
4$\sigma$-level, being consistent with contributions  
from both sources. Below 10 MeV only 3C~454.3 is significantly detected, with
the strongest evidence (5.6 $\sigma$) in the 3-10 MeV band. 
Significant flux variability is not observed for both sources, while 
a low emission is seen most of the years in the 3-10 MeV light curve for 3C~454.3. Its time-averaged MeV spectrum suggests a power maximum between 3 to 10 MeV.
\end{abstract}

\section*{Introduction}
Gamma-ray emission from  the 
two blazars 3C~454.3 and CTA~102, located $\sim$~7$^{\circ}$ apart
on the sky, was discovered by EGRET during the early Compton Gamma-Ray Observatory (CGRO) mission
\cite{Hartman93,Nolan93}. 
COMPTEL measurements at MeV energies of the same time periods, 
reported earlier by Blom et al. \cite{Blom95}, show indications that both sources are weakly 
detected in the COMPTEL uppermost (10-30 MeV) band.
If combined with EGRET, the trend for a spectral flattening at MeV energies 
becomes visible. 
At even lower energies (hard X-rays) OSSE detected variability of 
both sources in 1994 \cite{McNaron-Brown95}. 

In this paper, we present the results of the COMPTEL observations 
on both blazars during the whole CGRO mission. The data have been consistently analyzed for both blazars and compared to the results obtained early in the  mission, published by Blom et al. \cite{Blom95}. A more detailed presentation of the analysis results  will be given by Zhang et al. \cite{Zhang01}.

\section*{Data and Analysis Method}
The imaging Compton Telescope COMPTEL - one of four experiments aboard CGRO -  was sensitive to $\gamma$-rays in the energy range 0.75-30 MeV  \cite{Schonfelder93}. During its whole mission (1991 - 2000), the sources were  in 22 CGRO viewing period (VPs) within 40$^{\circ}$ of the COMPTEL pointing direction.  These VPs were selected in our analyses. 

The analyses were carried out by using the standard COMPTEL 
maximum-likelihood analysis procedure including a filtering technique 
for background generation. Point spread functions of the instrument which assume an E$^{-2}$ power law shape for the input spectrum were applied in our analyses.

\begin{figure}[t!] 
\psfig{file=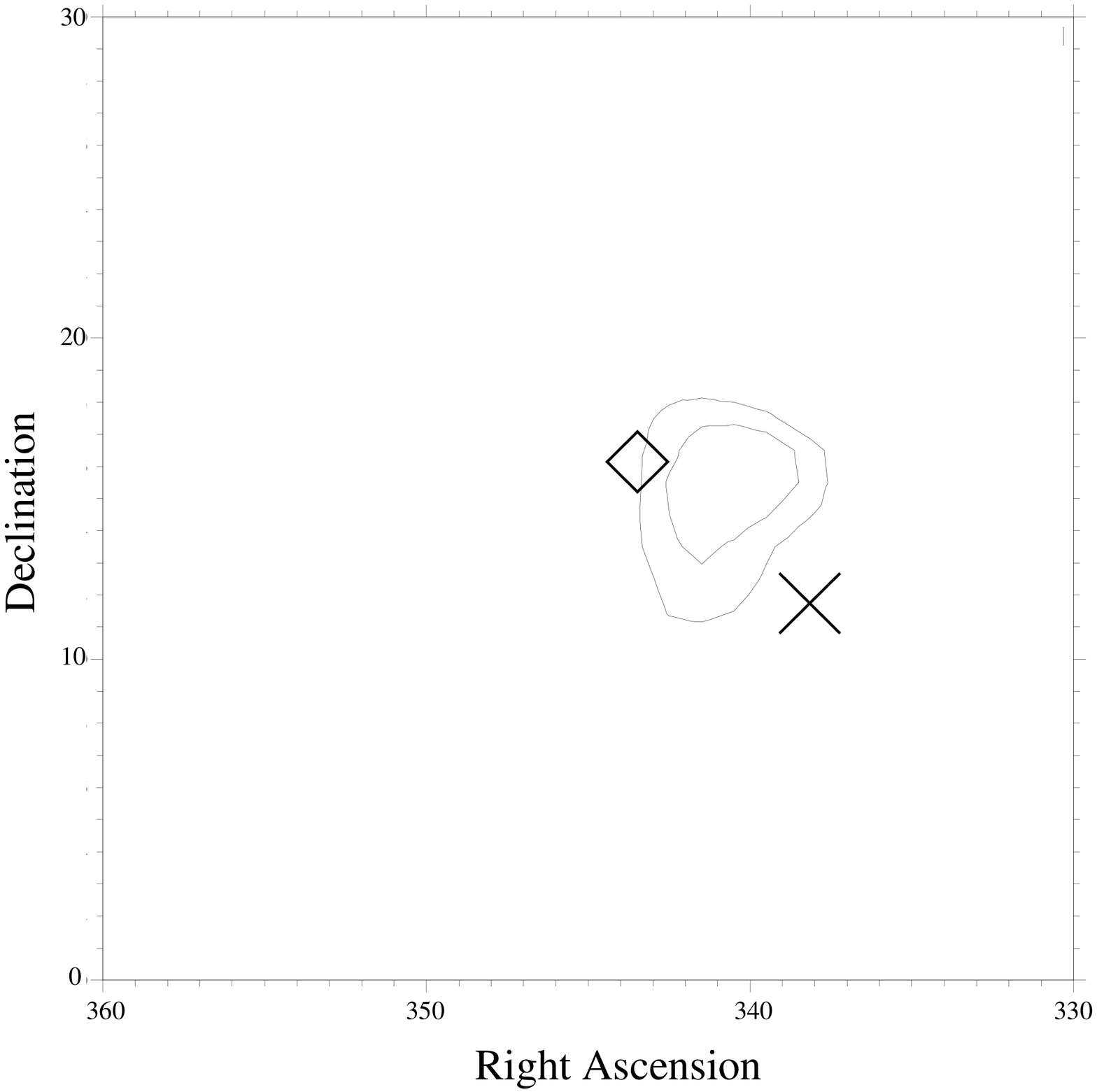,height=7.3cm,width=7.3cm}
\hfill
\epsfig{file=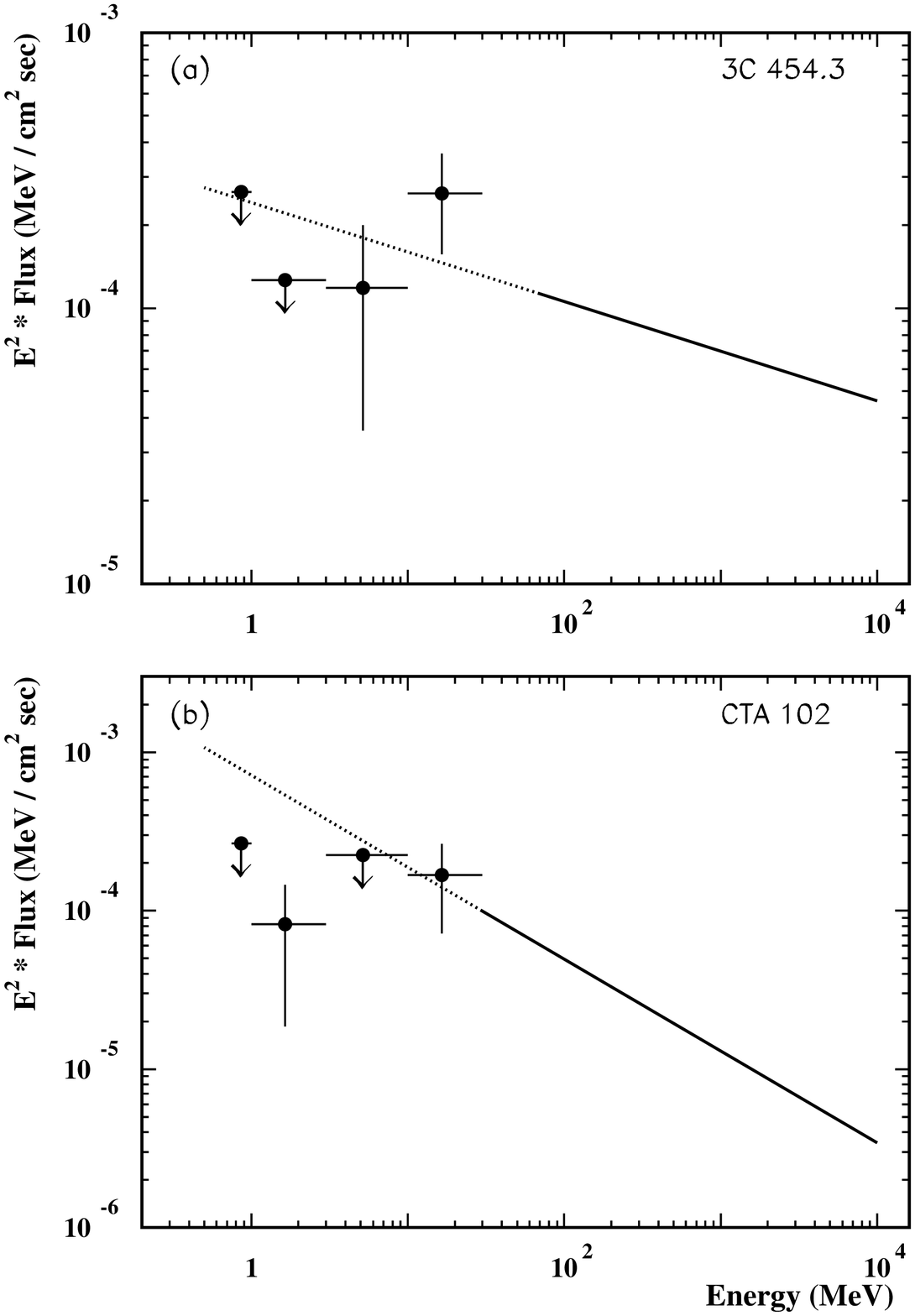,height=6.3cm,width=7.0cm}
\vspace{10pt}
\caption{
{\bf{Left:}} COMPTEL 10-30 MeV map for 3C~454.3 ($\diamond$)  and CTA~102 ($\times$) from Phase 1 observations in 1992.  The contour lines start at a detection significance level of 3 $\sigma$ with steps of 0.5 $\sigma$.
{\bf{Right:}} Combined COMPTEL/EGRET spectra for 3C~454.3 (a) and CTA~102 (b). The filled circles represent the COMPTEL spectra of Phase 1 (VPs 19.0, 26.0, 28.0 and 37.0), the solid lines the EGRET spectra of VP 19 [1, 2]  and the dotted lines their extrapolations toward lower energies. The error bars are 1 $\sigma$ and the upper limits 2 $\sigma$.}
\label{fig1}
\end{figure}

\section*{Results}
\subsection*{Phase 1}
For CGRO Phase 1 (April '91 to November '92) our results are similar to those reported by Blom et al. \cite{Blom95}. There is evidence for emission from the general direction of 3C~454.3 and CTA~102 in the 10-30 MeV band with a detection significance of $\sim$ 3.9 $\sigma$, 
which is consistent with contributions from both sources (Figure~\ref{fig1}). At lower energies, only marginal detections and upper limits are obtained.
The combined COMPTEL/EGRET spectra indicate spectral turnovers at MeV energies for both sources (Figure~\ref{fig1}). 
\subsection*{Phases 1-8}
\begin{figure}[t!] 
\psfig{file=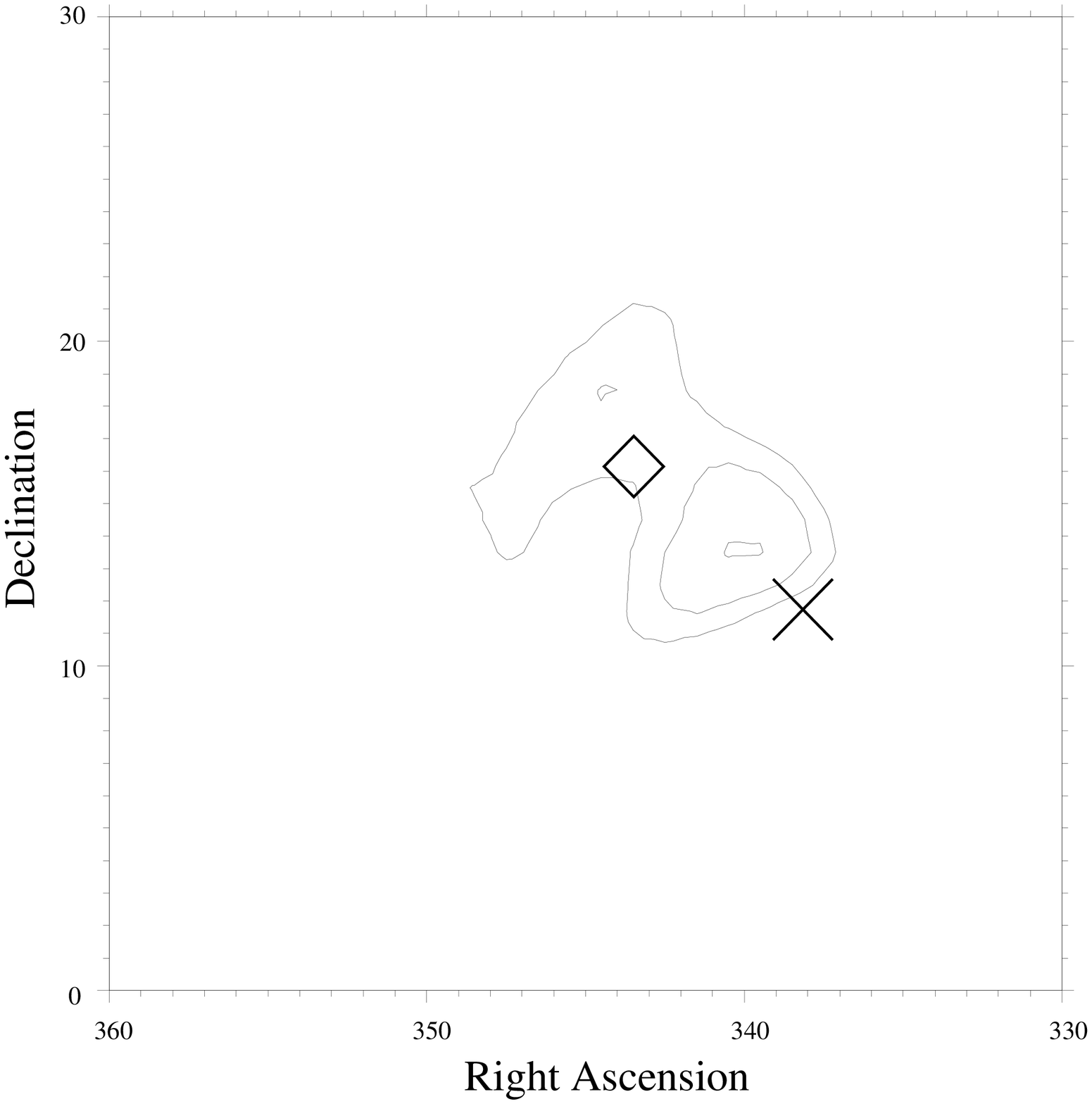,height=7.3cm,width=7.3cm}
\hfill
\psfig{file=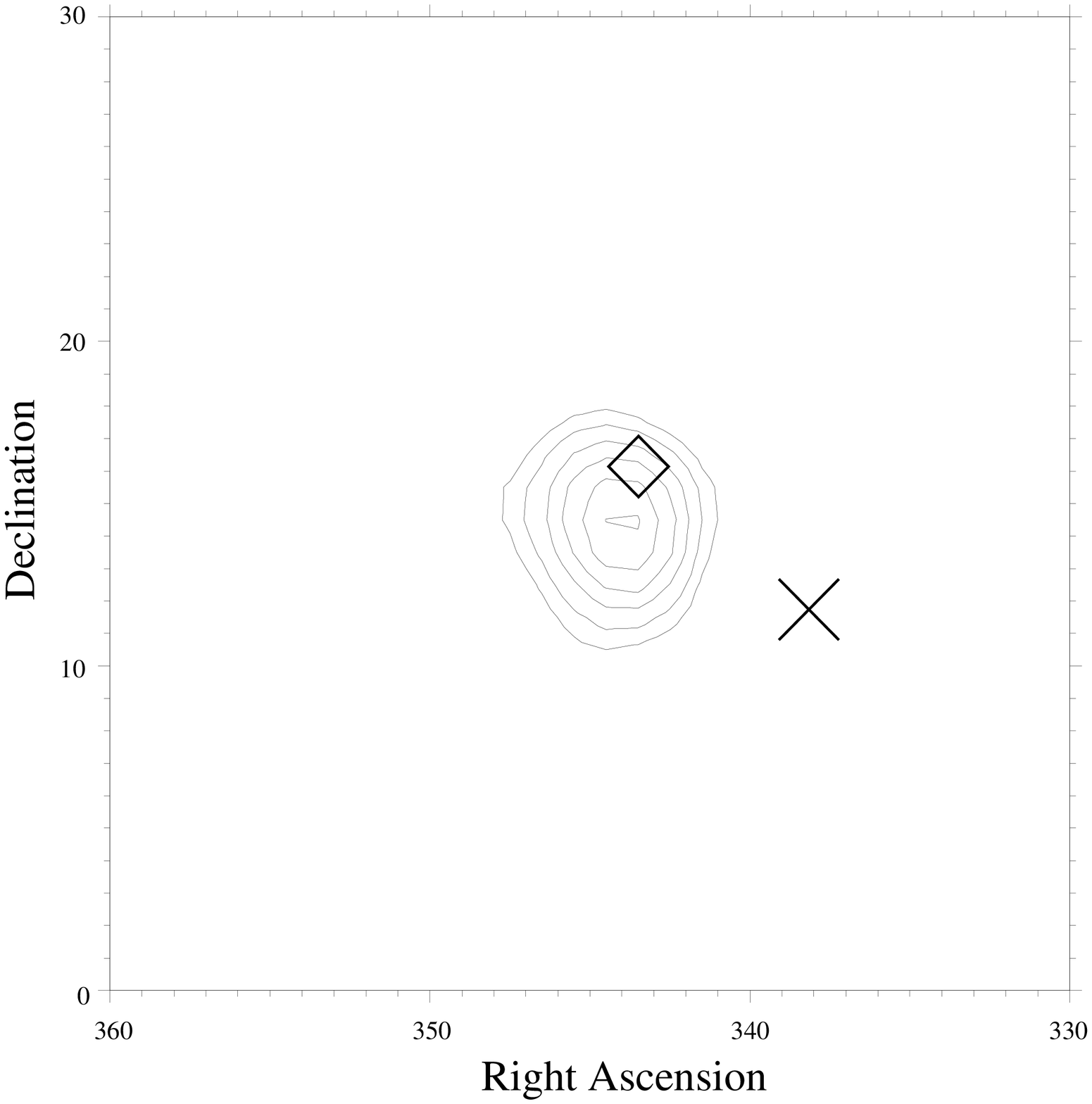,height=7.3cm,width=7.3cm}
\vspace{10pt}
\caption{
COMPTEL 10-30 MeV map (left) and 3-10 MeV map (right) for 3C~454.3 ($\diamond$)  and CTA~102 ($\times$) from 8 years of observations.  The contour lines start at a detection significance level of 3 $\sigma$ with steps of 0.5 $\sigma$. 
}
\label{fig2}
\end{figure}

No significant ($>$3$\sigma$) detection is found for either 3C~454.3
or CTA~102 in any individual COMPTEL observation during the whole CGRO mission in the four standard energy ranges.
However, combining all the COMPTEL data provides evidence for
3C~454.3 at energies above 1 MeV and for CTA~102 at energies above 10 MeV.
The 10-30 MeV skymap (Figure~\ref{fig2}) shows that a 4$\sigma$-excess is located
just between 3C~454.3 and CTA~102, again consistent with emission from both.
No signal can be derived from CTA~102 in the 3-10 MeV band, while an obvious 
excess is detected near 3C~454.3 at a significance level of 
$\sim$ 5.6 $\sigma$ (see Figure~\ref{fig2}). 3C~454.3 is located at the 4.5$\sigma$-detection contour and is within the 3 $\sigma$ error location contour. 
Since 3C~454.3 is  the only known $\gamma$-ray source in the COMPTEL error box,  we attribute this significant excess to emission from 3C~454.3.

\begin{figure}[t!] 
\epsfig{file=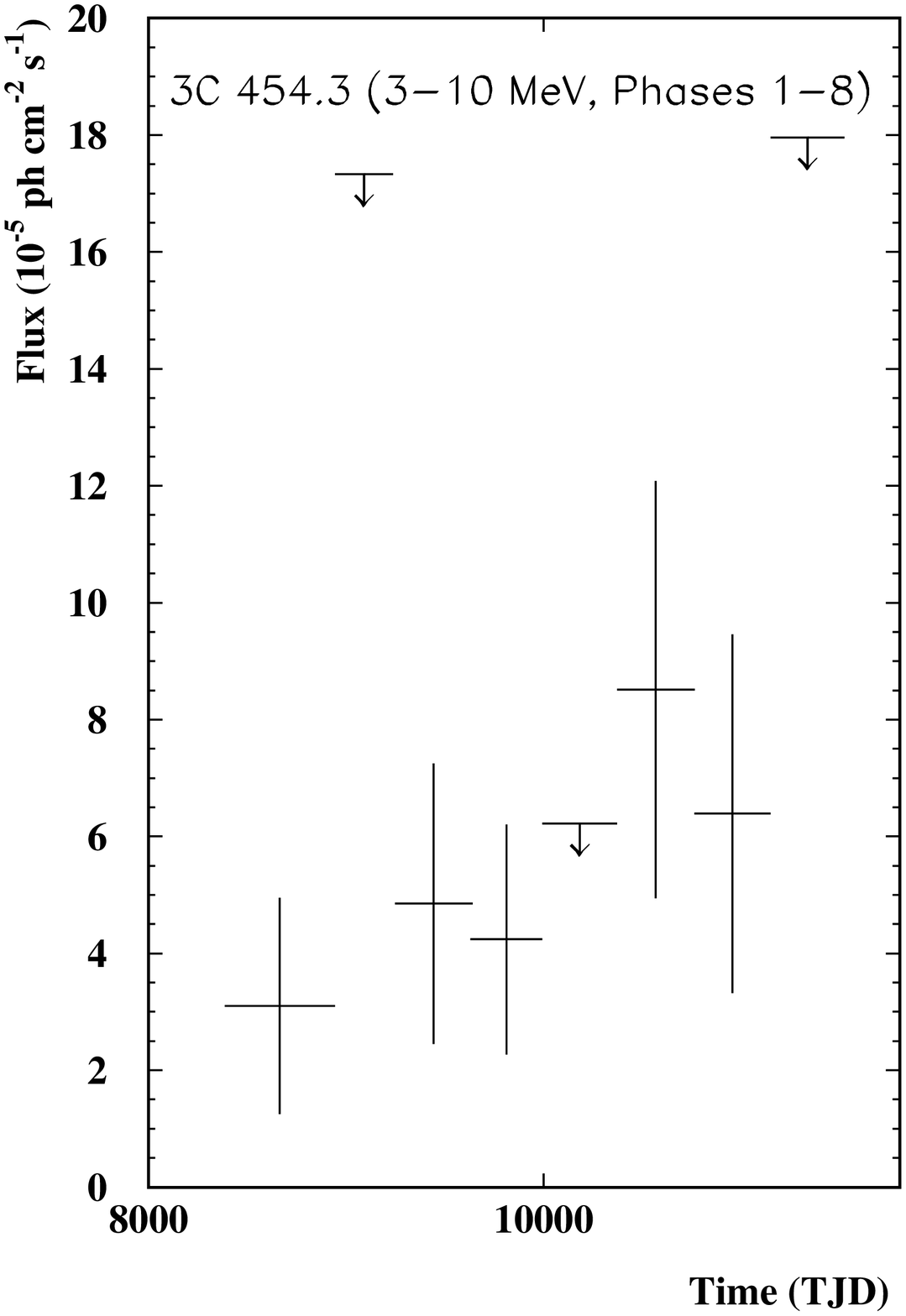,height=6.6cm,width=7.3cm}
\hfill
\epsfig{file=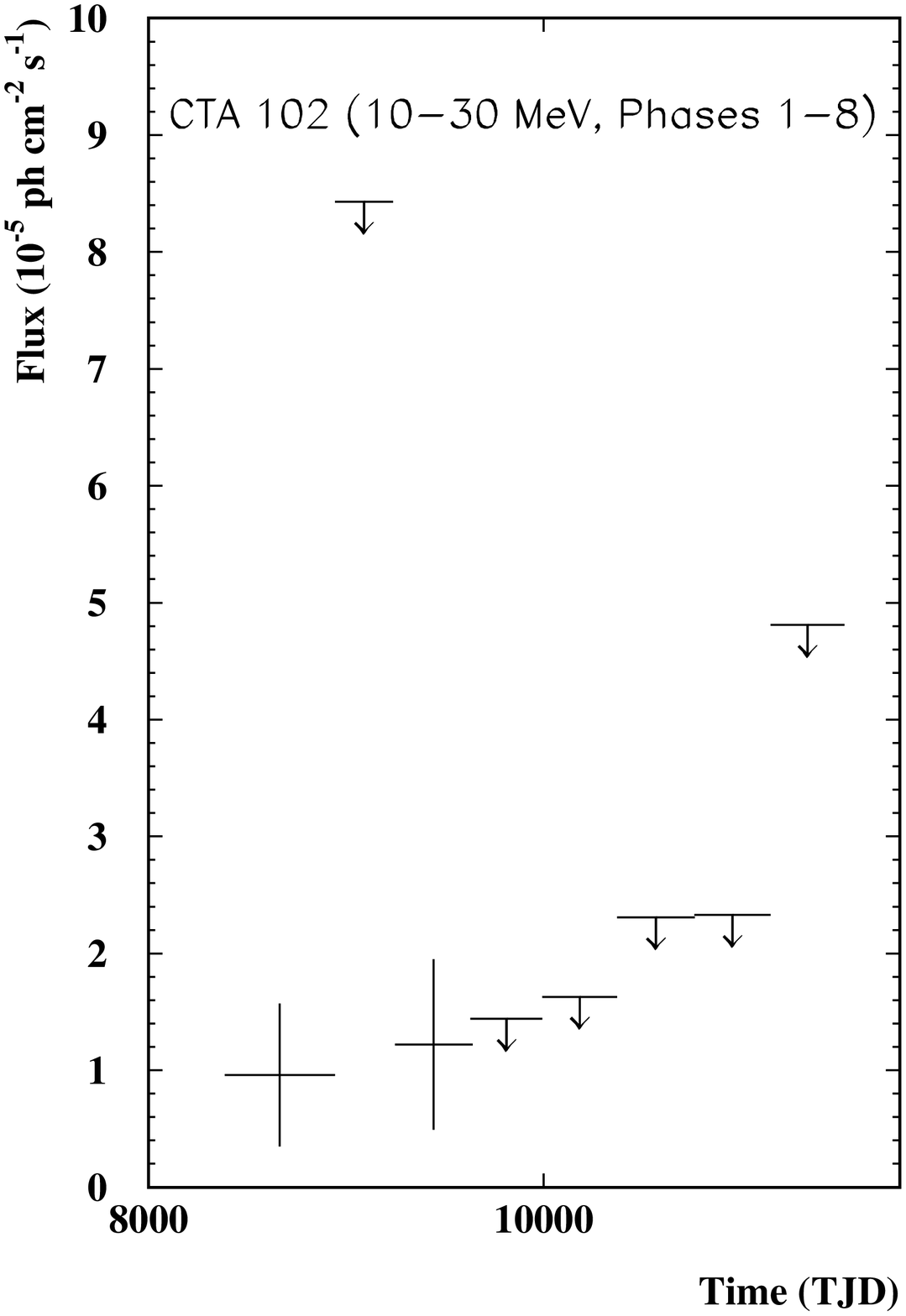,height=6.6cm,width=7.3cm}
\vspace{10pt}
\caption{
 The 3-10 MeV light curves for 3C~454.3 (left) and the 10-30 MeV light curve for CTA~102 (right) with  each bin averaged over one CGRO Phase which typically covers a time period of 1 year.
The high upper limits show times of low source exposures. The error bars are 1 $\sigma$ and the upper limits 2 $\sigma$.}
\label{fig3}
\end{figure}

A search for flux variability has been carried out in all individual
observations on 3C~454.3 and CTA~102 in the four standard COMPTEL energy bands.
All light curves are consistent with a constant flux for both sources 
over a period of about 8~years.
While only marginal hints or upper limits are found along the CGRO mission
 for CTA~102, 
3C~454.3 seem to be always emitting at somehow a low level in the
3-10~MeV band (Figure~\ref{fig3}). This is consistent with the significant detection of 3C~454.3 in the sum of the 3-10 MeV data. The 10-30 MeV light curve of CTA 102 (Figure~\ref{fig3}) shows mostly upper limits, indicating the source was weak along the CGRO mission.

\begin{figure}[t!] 
\epsfig{file=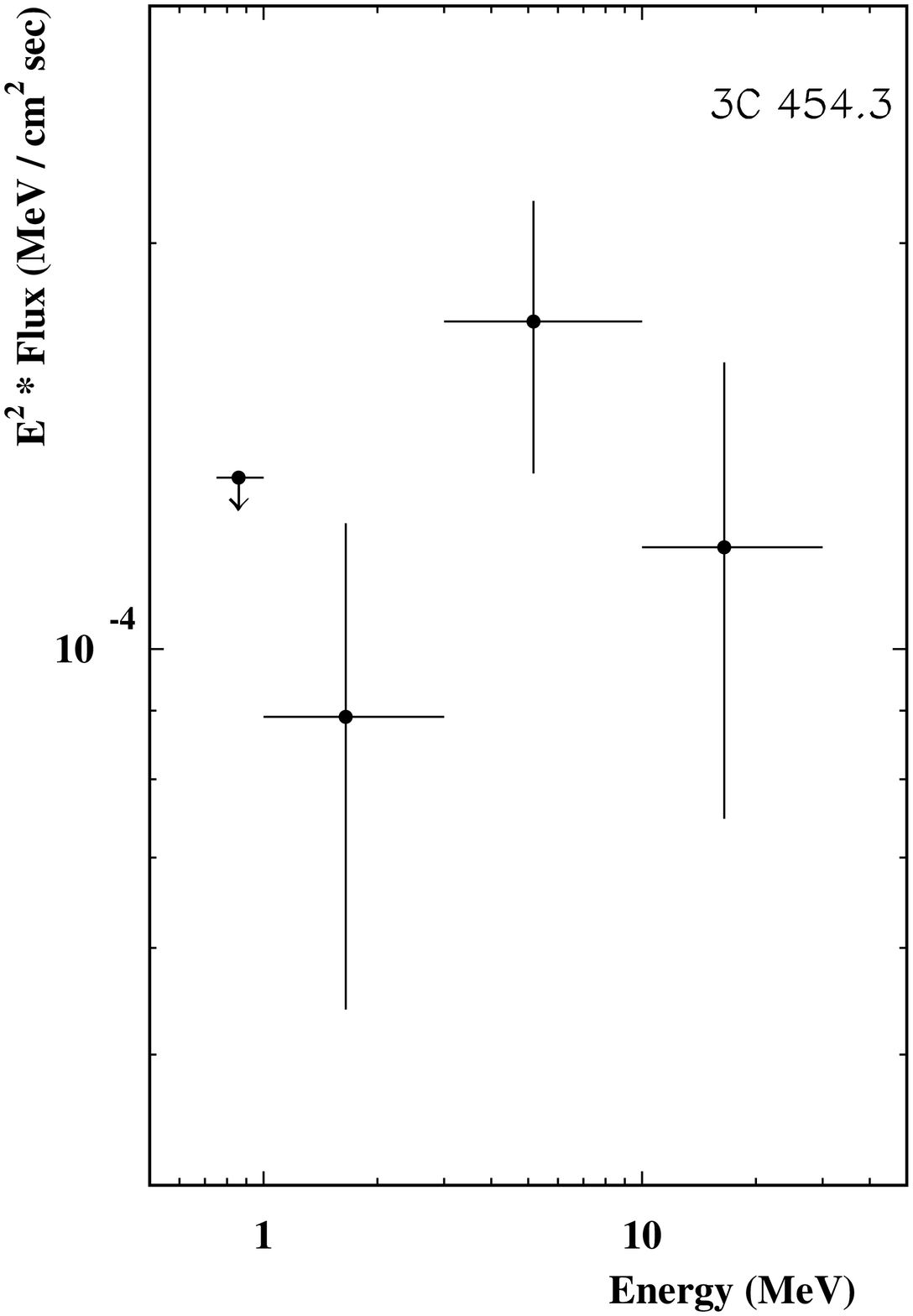,height=6.6cm,width=7.3cm}
\hfill
\epsfig{file=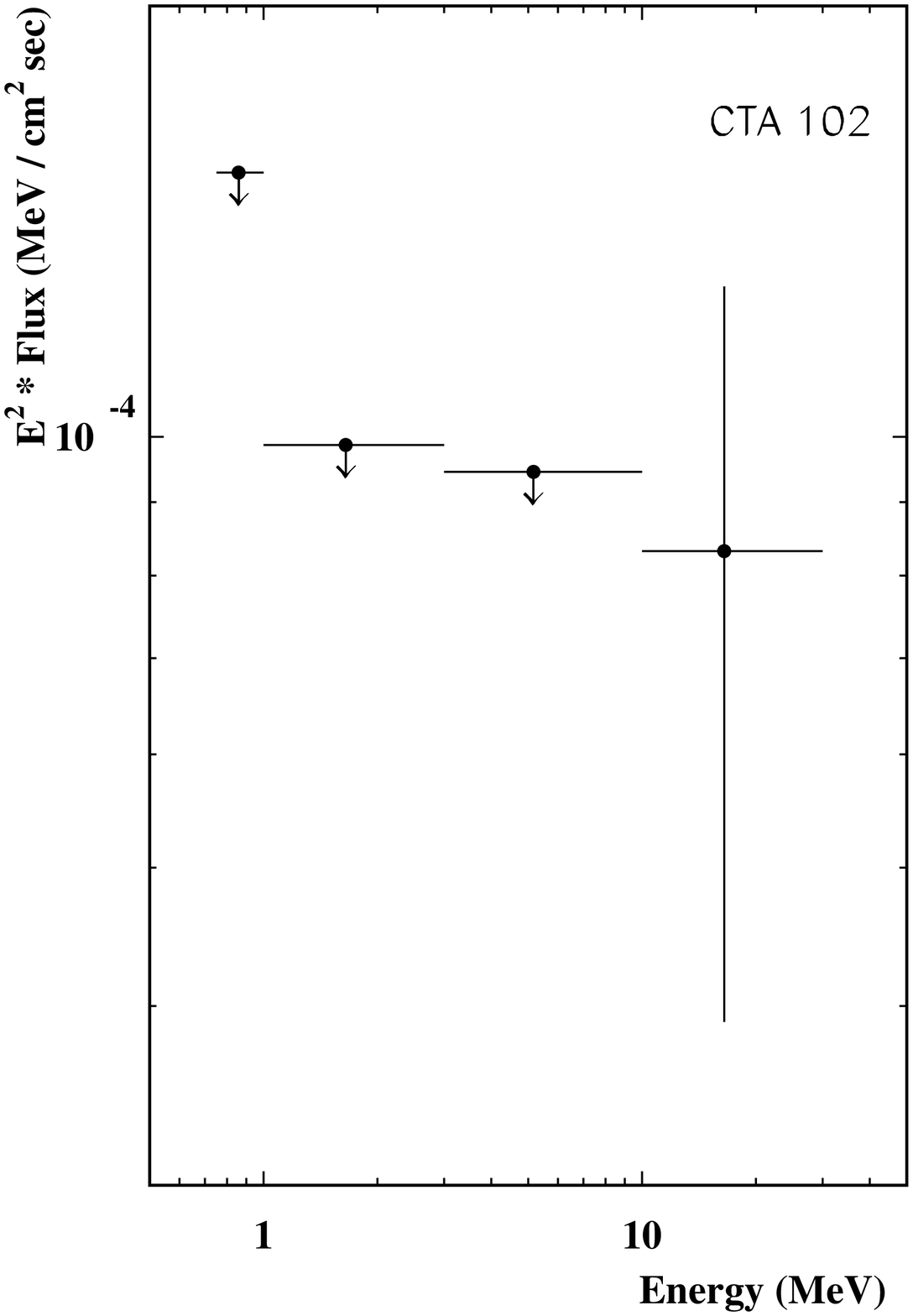,height=6.6cm,width=7.3cm}
\vspace{10pt}
\caption{
 COMPTEL spectra for 3C~454.3 (left) and CTA~102 (right) averaged over observations in Phases 1-8. The error bars are 1 $\sigma$ and the upper limits 2 $\sigma$.}
\label{fig4}
\end{figure}

Figure~\ref{fig4} shows the COMPTEL spectra  averaged over a time period between April 1991 and November 1999 for both sources.
The spectrum of 3C~454.3 (Figure~\ref{fig4}) suggests a power maximum (at least with respect to MeV energies) in the 3-10 MeV band. No conclusion can be derived from the spectrum of CTA 102 because only in the 10-30 MeV band a weak detection is obtained (Figure~\ref{fig4}).

\section*{Discussion and Summary}
Our analyses of the data of the early COMPTEL observations of the $\gamma$-loud flat-spectrum radio quasars  3C~454.3 and CTA~102 provide similar results with respect to the weak source detections and spectra as reported by Blom et al.\cite{Blom95}. 
The consistent analysis of all available COMPTEL data, covering 8 years,  reveals some new features for these two blazars. The most important one is the significant detection of 3C~454.3 in the 3-10 MeV band. Furthermore, 3C~454.3  seems to be a emitter always detectable in the 3-10 MeV band, where it appears to reach its power maximum in the time-averaged spectrum. It  resembles 3C~273 which is a rather stable MeV emitter with a power maximum in the 3-10 MeV band  \cite{Collmar01}, if the upper limits in the 3-10 MeV light curve are detections of 3C 454.3 near the threshold of COMPTEL. Therefore, 3C~454.3 might internally be similar to 3C~273 however at a lower flux level since we just see the 'tip of the iceberg' with COMPTEL.  Compared to 3C 454.3, the MeV emission from CTA 102 is weak and the detection significance marginal. 
Indications  for MeV emission from the source is found only in some early Phases. 
Its time-averaged MeV spectrum does not allow to draw any conclusions.

The COMPTEL results have to be compared to those in additional energy bands, in particular to the neighboring $\gamma$-ray bands covered by EGRET and OSSE. This work is in progress and will be given by Zhang et al. \cite{Zhang01}.

\vspace{0.4cm}\noindent 
{\small ACKNOWLEDGMENTS: The COMPTEL project is supported by the German government through DARA grant 50 QV 9096 8, by NASA under contract NAS5-26645, and by the Netherlands Organization for Scientific Research (NWO).}

\end{document}